\documentclass[aps,pra,twocolumn,nofootinbib]{revtex4}
\usepackage{amsmath, amsthm, amscd, amssymb}
\usepackage{bm}
\usepackage{bbm}
\usepackage{epsfig}
\usepackage{graphicx}
\usepackage{epstopdf}

\begin{document}

\title{GW170817 event rules out general relativity in favor of vector gravity%
}
\author{Anatoly A. Svidzinsky$^1$ and Robert C. Hilborn$^2$}
\affiliation{$^1$Department of Physics \& Astronomy, Texas A\&M University, College
Station, TX 77843 \\
$^2$American Association of Physics Teachers, One Physics Ellipse, College
Park, MD 20740 }
\date{\today }

\begin{abstract}
The observation of gravitational waves by the three LIGO-Virgo
interferometers allows the examination of the polarization of gravitational
waves. Here we analyze the binary neutron star event GW170817, whose source
location and distance are determined precisely by concurrent electromagnetic
observations. Applying a signal accumulation procedure to the LIGO-Virgo
strain data, we find ratios of the signals detected by the three
interferometers. We conclude that the signal ratios are inconsistent with
the predictions of general relativity, but consistent with the recently
proposed vector theory of gravity [Phys. Scr. 92, 125001 (2017)]. Moreover,
we find that vector gravity yields a distance to the source in agreement
with the astronomical observations. If our analysis is correct, Einstein's
general theory of relativity is ruled out in favor of vector gravity at $%
99\% $ confidence level and future gravitational wave detections by three or
more observatories should confirm this conclusion with higher precision.
\end{abstract}

\maketitle

\section{Introduction}

Recently, joint detection of gravitational waves by two LIGO interferometers
in the US and the Virgo interferometer in Italy became a reality \cite%
{Abbo17a,Abbo17b}. This achievement provides the opportunity to measure the
polarization of gravitational waves and, e.g., to determine whether gravity
is a pure tensor field, as predicted by general relativity \cite{Eins15}, or
a pure vector field described, for example, by the vector theory of gravity 
\cite{Svid17,Svid18}. It has long been realized \cite{Eard73,Eard73a} that
determining the tensor versus vector nature of gravitational wave
polarization provides a critical test of general relativity. In fact, as C.
Will has noted \cite{Will14} \textquotedblleft If distinct evidence were
found of any mode other than the two transverse quadrupolar modes of general
relativity, the result would be disastrous for general
relativity.\textquotedblright

Einstein's general relativity is an elegant theory which postulates that
space-time geometry itself (as embodied in the metric tensor) is a dynamical
gravitational field. However, the beauty of the theory does not guarantee
that the theory describes nature. Although it is remarkable that general
relativity, born more than 100 years ago, has managed to pass many
unambiguous observational and experimental tests, it has undesirable
features. For example, general relativity is not compatible with quantum
mechanics and it can not explain the value of the cosmological term (dark
energy), to name a few.

Recently, a new alternative vector theory of gravity was proposed \cite%
{Svid17,Svid18}. The theory assumes that gravity is a vector field in fixed
four-dimensional Euclidean space which effectively alters the space-time
geometry of the Universe. The direction of the vector gravitational field
gives the time coordinate, while perpendicular directions are spatial
coordinates. Similarly to general relativity, vector gravity postulates that
the gravitational field is coupled to matter through a metric tensor which
is, however, not an independent variable but rather a functional of the
vector gravitational field.

Despite fundamental differences, vector gravity also passes all available
gravitational tests \cite{Svid17}. In addition, vector gravity provides an
explanation of dark energy as the energy of the longitudinal gravitational
field induced by the expansion of the Universe and yields, with no free
parameters, the value of $\Omega _{\Lambda }=2/3$ \cite{Svid18} which agrees
with the results of Planck collaboration \cite{Planck14} and recent results
of the Dark Energy Survey. Thus, vector gravity solves the dark energy
problem.

In order to determine whether the gravitational field has a vector or a
tensor character, additional tests are required. Here we conduct such a test
based on gravitational-wave strain data released by the LIGO-Virgo
collaboration for the GW170817 event \cite{LV17}. We find that predictions
of vector gravity are compatible with these data. The predictions of general
relativity are not compatible with the data and hence, general relativity is
ruled out (at 99\% confidence level). Our conclusion is opposite to that of
the LIGO-Virgo collaboration \cite{Max18}. In Section \ref{Comment} we show
why we believe the LIGO-Virgo analysis underestimates the LIGO-Livingston
strain signal. For the GW170817 source location, that underestimate leads to
the erroneous conclusion that the GW170817 data strongly favor tensor
polarization of gravitational waves over vector polarization.

Both in general relativity and vector gravity the polarization of
gravitational waves emitted by orbiting binary objects is transverse, that
is, a gravitational wave (GW) yields motion of test particles in the plane
perpendicular to the direction of wave propagation. However, the response of
the laser interferometer for the GW is different in the two theories \cite%
{Nish09,Chat12,Haya13,Pitk17,Alle18,Take18,Hagi18}. This difference can be
used to test the theories. In particular, previous work \cite%
{Nish09,Chat12,Haya13,Pitk17,Alle18,Take18,Hagi18} shows that if the GW
source location is precisely known and if the relative (complex) amplitudes
among the three observatories are measured with sufficient precision, then
the polarization character of the GWs can be found by relatively
straightforward means.

For a weak transverse gravitational wave propagating along the $x-$axis in
vector gravity the equivalent metric evolves as \cite{Svid17}%
\begin{equation}
g_{ik}=\eta _{ik}+\left( 
\begin{array}{cccc}
0 & 0 & h_{0y}(t,x) & h_{0z}(t,x) \\ 
0 & 0 & 0 & 0 \\ 
h_{0y}(t,x) & 0 & 0 & 0 \\ 
h_{0z}(t,x) & 0 & 0 & 0%
\end{array}%
\right) ,  \label{met1}
\end{equation}%
where $\eta _{ik}$ is the Minkowski metric. The three-dimensional
gravitational field vector of this wave is 
\begin{equation}
\mathbf{h}=(0,h_{0y},h_{0z}).
\end{equation}

\begin{figure}[h]
\centering
\hspace*{-0.5cm} \includegraphics[width=7cm,angle=270,origin=c]{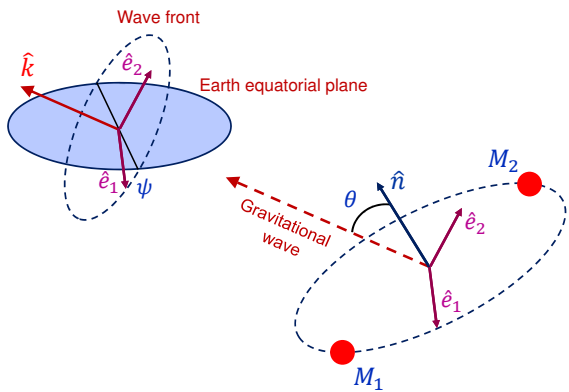}
\par
\vspace{-1.4cm}
\caption{Two stars are orbiting each other and emit gravitational waves
which are detected on Earth. Orientation of the orbital plane is described
by a unit vector $\hat{n}$ perpendicular to the plane. The angle between $%
\hat{n} $ and the direction of the wave propagation $\hat{k}$ is the orbit
inclination angle $\protect\theta $. The wave polarization vector $\hat{e}%
_{1}$ is perpendicular to $\hat{k}$ and chosen to be parallel to the orbital
plane, while $\hat{e}_{2}=\hat{k}\times \hat{e}_{1}$. The polarization angle 
$\protect\psi $ is the angle between $\hat{e}_{1}$ and the line formed by
intersection of the wave front and Earth's equatorial plane.}
\label{Fig1}
\end{figure}

We will consider a general case of a GW propagating along the unit vector $%
\hat{k}$; the vector $\mathbf{h}$ is perpendicular to $\hat{k}$. To be
specific, we consider the generation of GWs by two compact stars with masses 
$M_{1}$ and $M_{2}$ moving along circular orbits with angular velocity $%
\Omega (t)$. We assume that the spacing between the stars is much larger
than their dimensions and the motion is non-relativistic. For example, the
spacing between two neutron stars of equal masses $1$ $M_{\odot }$ moving
with orbital frequency $50$ Hz (i.e. the GW frequency is $100$ Hz) is $r=140$
km, which is much larger than the stellar radii $\sim 10$ km. The orbital
velocity of the stars is $0.07c$. For such parameters the orbital frequency
and the radius change due to emission of GWs only a little during an orbital
period and one can use an adiabatic approximation. This example is relevant
for the GW170817 event signal at the early inspiral stage which we analyze
in this paper. At this stage, the energy loss by the binary system is
described by the same quadrupole formula in vector gravity and general
relativity, and, hence, both theories yield the same gravitational waveform.
Such a waveform is known analytically and is accurate during almost the
entire data collection time interval (apart from the last second or so
before merger) for the GW170817 event involving low-mass objects.

In contrast to circular-orbit binaries, the velocity on an eccentric orbit
changes over its period and the instantaneous orbital frequency also varies
substantially. Orbital eccentricity leads to multiple orbital frequency
harmonics in the gravitational waveform. However, since GW emission tends to
circularize the orbit as it shrinks the binary separation \cite{Pete64},
many of the binary GW sources are expected to have small orbital
eccentricity by the time they enter the frequency bands of ground-based GW
detectors. For example, although the eccentricity of the Hulse-Taylor binary
pulsar system is currently $0.6171$ \cite{Huls75} it will have an
eccentricity of $\sim 10^{-4}$ when it enters the LIGO band \cite{Gond17}.
Orbital eccentricity was assumed to be zero in the analysis of the detected
GW sources \cite{Abbo17b,Abbo18} and we make the same assumption in the
present paper.

Neutron stars spin down due to the loss of rotational energy by powering
magnetically driven plasma winds. As a consequence, the spin of neutron
stars in isolated binary systems is expected to be relatively small by the
time of their merger. This is consistent with the spin of neutron stars
measured in the Galactic binaries (see Table III in \cite{Zhu18}). Since no
evidence for non-zero component spins was found for the binary neutron star
merger GW170817 \cite{Abbo18} we will disregard effects of the neutron star
spins in the present analysis.

We denote unit polarization basis vectors as $\mathbf{\hat{e}}_{1}$ and $%
\mathbf{\hat{e}}_{2}$. They are perpendicular to $\hat{k}$ and we choose
them as shown in Fig. \ref{Fig1}. For non-relativistic motion, using
formulas of Ref. \cite{Svid17}, we obtain for vector GW in the adiabatic
approximation far from the binary system 
\begin{equation}
\mathbf{h}=A\left[ \sin \theta \sin (2\phi )\mathbf{\hat{e}}_{1}+\frac{1}{2}%
\sin (2\theta )\cos (2\phi )\mathbf{\hat{e}}_{2}\right] ,  \label{s1}
\end{equation}%
where 
\begin{equation}
A=\frac{4GM\Omega ^{2}r^{2}}{c^{4}R}=\frac{5^{1/4}G^{5/4}M^{5/3}}{%
c^{11/4}RM_{c}^{5/12}}\frac{1}{\left( t_{c}-t\right) ^{1/4}}  \label{aa}
\end{equation}%
is the GW amplitude which gradually increases with time due to the increase
of the orbital velocity of the stars caused by emission of GWs. In Eqs. (\ref%
{s1}) and (\ref{aa}) $M=M_{1}M_{2}/(M_{1}+M_{2})$ is the reduced stellar
mass, $r=r(t)$ is the distance between the stars, 
\begin{equation}
\phi (t)=-\left( \frac{c^{3}}{5GM_{c}}\right) ^{5/8}\left( t_{c}-t\right)
^{5/8}+\phi _{0}  \label{aa0}
\end{equation}%
is the star azimuthal angle in the orbital plane, $t_{c}$ is the coalescence
time, 
\begin{equation}
M_{c}=\frac{\left( M_{1}M_{2}\right) ^{3/5}}{(M_{1}+M_{2})^{1/5}}  \label{mc}
\end{equation}%
is the chirp mass of the system, $R$ is the distance to the binary system
and $\theta $ is the orbit inclination angle, that is, the angle between the
normal to the orbital plane $\hat{n}$ and the direction of the wave
propagation $\hat{k}$ (see Fig. \ref{Fig1}). Depending on the inclination
angle of the orbital plane of the binary stars, the GW in vector gravity can
be linearly or elliptically polarized in the same way as electromagnetic
waves generated by an oscillating quadrupole.

The signal of the LIGO-like interferometer with perpendicular arms of length 
$L_{a}$ along the direction of unit vectors $\hat{a}$ and $\hat{b}$ is
proportional to the relative phase shift of the laser beam traveling a
roundtrip distance $2L_{a}$ along the two arms. The relative phase shift
divided by $2L_{a}\omega /c$, where $\omega $ is the angular frequency of
the optical field in the interferometer, gives the gravitational wave strain 
$h(t)$ \cite{Svid17,Nish09,Chat12,Haya13,Pitk17,Alle18} 
\begin{equation}
h(t)=(\hat{a}\cdot \hat{k})(\hat{a}\cdot \mathbf{h})-(\hat{b}\cdot \hat{k})(%
\hat{b}\cdot \mathbf{h}).  \label{wa4}
\end{equation}%
Using Eqs. (\ref{s1}) and (\ref{wa4}) yields the following expression for
the interferometer response for the vector GW 
\begin{equation}
h(t)=A\left( \sin \theta \sin (2\phi )V_{1}+\frac{1}{2}\sin (2\theta )\cos
(2\phi )V_{2}\right) ,  \label{s4}
\end{equation}%
where $V_{1}$ and $V_{2}$ are the detector response functions for the two
basis vector polarizations $\mathbf{\hat{e}}_{1}$ and $\mathbf{\hat{e}}_{2}$ 
\begin{equation}
V_{1,2}=(\hat{a}\cdot \hat{k})(\hat{a}\cdot \mathbf{\hat{e}}_{1,2})-(\hat{b}%
\cdot \hat{k})(\hat{b}\cdot \mathbf{\hat{e}}_{1,2}).  \label{s5}
\end{equation}

If the same binary system emits tensor GWs according to general relativity,
the metric oscillates in the $\mathbf{\hat{e}}_{1}-\mathbf{\hat{e}}_{2}$
plane as

\begin{equation*}
h_{12}=h_{21}=A\cos \theta \sin (2\phi ),
\end{equation*}%
\begin{equation*}
h_{11}=-h_{22}=\frac{A}{2}\left( 1+\cos ^{2}\theta \right) \cos (2\phi ),
\end{equation*}%
and the interferometer response is given by%
\begin{equation}
h(t)=A\left( \frac{1}{4}\left( 1+\cos ^{2}\theta \right) \cos (2\phi
)T_{1}+\cos \theta \sin (2\phi )T_{2}\right) ,  \label{wa5}
\end{equation}%
where $T_{1,2}$ are the interferometer response functions for the two basis
tensor polarizations 
\begin{equation}
T_{1}=(\hat{a}\cdot \mathbf{\hat{e}}_{1})^{2}-(\hat{b}\cdot \mathbf{\hat{e}}%
_{1})^{2}+(\hat{b}\cdot \mathbf{\hat{e}}_{2})^{2}-(\hat{a}\cdot \mathbf{\hat{%
e}}_{2})^{2},  \label{s6}
\end{equation}%
\begin{equation}
T_{2}=(\hat{a}\cdot \mathbf{\hat{e}}_{1})(\hat{a}\cdot \mathbf{\hat{e}}%
_{2})-(\hat{b}\cdot \mathbf{\hat{e}}_{1})(\hat{b}\cdot \mathbf{\hat{e}}_{2}),
\label{s7}
\end{equation}%
and $A(t)$, $\phi (t)$ are given by the same Eqs. (\ref{aa}) and (\ref{aa0})
as for vector gravity.

Equations (\ref{s4}) and (\ref{wa5}) show that vector GWs emitted parallel
to the orbital plane ($\theta =90^{0}$) will produce the same maximum
detector response as the general relativistic GWs emitted in the direction
perpendicular to the plane ($\theta =0^{0},$ $180^{0}$).

Next we introduce an integrated complex interferometer response 
\begin{equation}
I(t)=\int_{t_{0}}^{t_{0}+t}\left( t_{c}-\tau \right) ^{1/4}e^{-2i\phi (\tau
)}h(\tau )d\tau ,  \label{x3}
\end{equation}%
where $t$ is the signal collection time and $h(t)$ is the strain measured by
the interferometer that contains both signal and noise. According to Eqs. (%
\ref{aa}), (\ref{s4}) and (\ref{wa5}), the signal contribution to $I(t)$ is
proportional to $t$ provided we disregard small correction produced by the
fast oscillating term. Thus, signal accumulates with an increase of the
collection time $t$. In contrast, noise does not accumulate with $t$ and for
large enough $t$ the noise contribution to $I(t)$ can be disregarded. In
this case for vector gravity we obtain 
\begin{equation}
I_{V}(t)=\alpha \left( -i\sin (\theta )V_{1}+\frac{1}{2}\sin (2\theta
)V_{2}\right) t,  \label{f1a}
\end{equation}%
while for general relativity 
\begin{equation}
I_{T}(t)=\alpha \left( \frac{1}{4}\left( 1+\cos ^{2}\theta \right)
T_{1}-i\cos (\theta )T_{2}\right) t,  \label{f2a}
\end{equation}%
where 
\begin{equation}
\alpha =\frac{5^{1/4}G^{5/4}M^{5/3}}{2c^{11/4}RM_{c}^{5/12}}
\end{equation}%
is independent of time and the orientation of the interferometer arms.

Equations (\ref{f1a}) and (\ref{f2a}) are predictions of vector gravity and
general relativity which are valid with high accuracy if the orbital
frequency and radius slowly change during an orbital period. Equations (\ref%
{f1a}) and (\ref{f2a}) show that the amplitude of $I(t)$ grows linearly with
the collection time $t$ and the phase of $I(t)$ is independent of $t$. Both
the phase and the amplitude of $I(t)$ depend on the interferometer
orientation for the vector and tensor polarizations in a fixed way. This
paper tests this dependence. A verified discrepancy between observations and
the predictions of a theory rules out the theory.

According to Eqs. (\ref{f1a}) and (\ref{f2a}), the ratio of the accumulated
signals $I(t)$ measured by two interferometers, e.g. LIGO-Hanford and
LIGO-Livingston, is a complex number. We denote this ratio as $H/L$. Vector
gravity predicts that%
\begin{equation}
\frac{H}{L}=\frac{2\sin (\theta )V_{H1}+i\sin (2\theta )V_{H2}}{2\sin
(\theta )V_{L1}+i\sin (2\theta )V_{L2}},  \label{f10}
\end{equation}%
while for the tensor polarization of general relativity%
\begin{equation}
\frac{H}{L}=\frac{\left( 1+\cos ^{2}\theta \right) T_{H1}-4i\cos (\theta
)T_{H2}}{\left( 1+\cos ^{2}\theta \right) T_{L1}-4i\cos (\theta )T_{L2}}.
\label{f11}
\end{equation}%
The complex ratio $H/L$ depends on the wave polarization (tensor or vector),
the direction of the wave propagation $\hat{k}$, the orientation of the
interferometer arms, the orbit inclination angle $\theta $ and the
polarization orientation angle $\psi $ defined in Fig. \ref{Fig1}.

Taking the Fourier transform of Eqs. (\ref{s4}) and (\ref{wa5}), we obtain
the interferometer signal in the frequency representation 
\begin{equation}
h_{V}(f)=B(f)\left[ -i\sin (\theta )V_{1}+\frac{1}{2}\sin (2\theta )V_{2}%
\right]  \label{f1}
\end{equation}%
for vector gravity, and%
\begin{equation}
h_{T}(f)=B(f)\left[ \frac{1}{4}\left( 1+\cos ^{2}\theta \right) T_{1}-i\cos
(\theta )T_{2}\right]  \label{f2}
\end{equation}%
for general relativity. Thus, in the frequency representation the ratio of
signals $h(f)$ measured by two interferometers is also given by Eqs. (\ref%
{f10}) and (\ref{f11}). The Fourier representation can be used to check the
correctness of the signal ratios obtained by the signal accumulation
approach.

For the LIGO-Virgo network, the orientation of the interferometer arms and
interferometer positions are accurately known \cite{LV16}. If an optical
counterpart of the GW source is found then the propagation direction of GW
is also accurately known. In this case the orbit inclination angle $\theta $
and polarization angle $\psi $ are the only free parameters in the signal
ratios.

The three interferometers of the LIGO-Virgo network yield two independent
complex ratios $H/L$ and $V/L$. In an ideal case of strong signals these two
complex numbers can be accurately measured and compared with the values
predicted by general relativity and vector gravity. The two measured complex
numbers $H/L$ and $V/L$ depend on two real variables $\theta $ and $\psi $.
Such a system of four equations with\ two unknowns is overdetermined and in
the general case can have a solution only for tensor or vector GWs, but not
for both. Thus, GW signals detected by a network of three interferometers
can, in principle, decide between tensor polarization (general relativity)
and vector polarization (vector gravity).

In the literature it is often mentioned that the arms of the two LIGO
interferometers (Livingston and Hanford) are almost co-aligned, and hence,
the ratio $H/L$ cannot give any information on the GW polarization. In fact,
the angles between the corresponding arms of the two LIGO interferometers
are actually not that small (taking the dot product between the arm
directions yields angles $13^{\circ }$ and $24^{\circ }$ respectively, and $%
27^{\circ }$ between the normals to the detector planes) and hence the $H/L$
ratio can impose significant constraints on the GW polarization detected by
the LIGO-Virgo network \cite{Hilb18}.

The LIGO-Virgo detection of the GW170814 event \cite{Abbo17a}, attributed to
binary black holes, presented the first possibility of testing the
polarization properties of GWs. However, for that event, there were no
concurrent electromagnetic observations, so the source location precision,
although much better than the previous two-observatory detections, was not
sufficient to decide between tensor and vector polarizations \cite%
{Max18,Hilb18}. The full parameter estimate of Ref. \cite{Abbo17a}
constrained the position of the GW source to a $90\%$ credible area of $60$
deg$^{2}$. It has been shown that vector GW polarization is compatible with
the GW170814 event in a considerable part of the $90\%$ credible area \cite%
{Hilb18}. As shown explicitly in Ref. \cite{Hilb18}, the uncertainty in the
GW170814 sky source location does not permit drawing definite conclusions
about the GW polarization. The same would be true for GW170817 (as confirmed
in Ref. \cite{Max18}, p. 12) absent the precise location information from
the concurrent detection of electromagnetic emissions.

On 17 August 2017, the Advanced LIGO and Virgo detectors observed the
gravitational wave event GW170817 - a strong signal from the merger of a
binary neutron-star system \cite{Abbo17b}. For the GW170817 event the
propagation direction of the GW is accurately known from the precise
coordinates of the optical counterpart discovered in close proximity to the
galaxy NGC 4993. Namely, at the moment of detection the source was located
at the latitude 23.37$^{\circ }$ S and longitude 40.8$^{\circ }$ E.

It turns out that for this source location we can distinguish between the
predictions of general relativity and vector gravity even if the ratios $H/L$
and $V/L$ are obtained with relatively poor accuracy. In the next section we
extract those ratios from the GW strain data released by the LIGO-Virgo
collaboration \cite{LV17}. In Section \ref{Test} we test vector gravity and
general relativity based on the obtained ratios. In Section \ref{Comment} we
explain why polarization analysis of GW170817 by the LIGO-Virgo
collaboration failed to find inconsistency with general relativity.

\section{GW170817 event: Data processing and signal extraction}

\label{Data}

As we have shown in the previous section, the crucial quantities for our
analysis are the complex amplitude ratios $H/L$ and $V/L$. Here we use
published strain time series for the three detectors \cite{LV17}. These data
are not normalized to the detector's noise (\textquotedblleft
whitened\textquotedblright ) and, thus, can be directly used to estimate the
GW signal ratios $H/L$ and $V/L$.

We bandpassed the time series data between $40$ Hz and $250$ Hz to eliminate
low and high frequency noise, which improves signal processing. A short
instrumental noise transient appeared in the LIGO-Livingston detector $1.1$
s before the coalescence time of GW170817. We restrict our analysis to the
prior time in order to avoid this \textquotedblleft glitch\textquotedblright
.

\begin{figure}[h]
\centering
\hspace*{-1cm} \includegraphics[width=13cm]{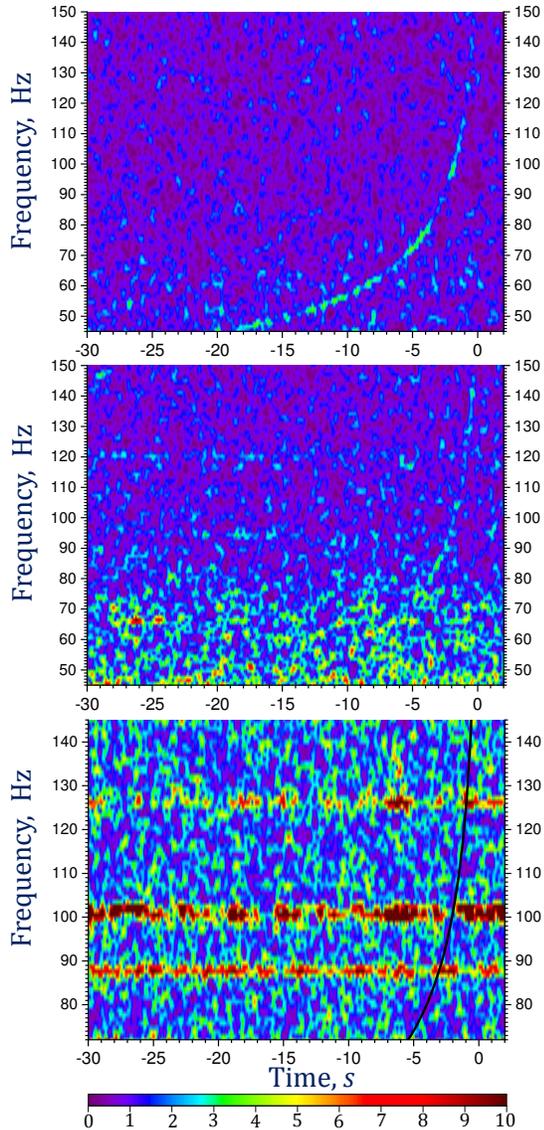}
\par
\vspace{-0.8cm}
\caption{Spectrograms (time-frequency representations) of strain data
containing the gravitational-wave event GW170817, observed by the
LIGO-Livingston (top), LIGO-Hanford (middle), and Virgo (bottom) detectors.
As in Ref. \protect\cite{Abbo17b}, times are shown relative to August 17,
2017 12:41:04 UTC. The amplitude scale in each spectrogram is the same. The
expected position of the Virgo signal is indicated as a black solid line in
the Virgo spectrogram.}
\label{Fig2}
\end{figure}

Using the GW170817 source location information \cite{Abbo17b} we calculated
the arrival time delays of the GWs at the interferometer locations. We find
that the GW arrived at the Virgo detector $0.02187$ s earlier than at the
LIGO-Livingston location, and $0.00333$ s later at the LIGO-Hanford
detector. We adjusted the measured strain time series for these time delays.

In Fig. \ref{Fig2} we plot spectrograms (time-frequency representations) of
strain data containing the gravitational-wave event GW170817, observed by
the LIGO-Livingston (top), LIGO-Hanford (middle), and Virgo (bottom)
detectors. As in Ref. \cite{Abbo17b}, times are shown relative to August 17,
2017 12:41:04 UTC. The amplitude scale in each detector is the same. In
contrast to Ref. \cite{Abbo17b}, we have not normalized the data to the
detector's noise amplitude spectral density which allows us to estimate the
ratios of the signal amplitudes in different detectors. The figure shows
that the GW signal is visible in the LIGO-Livingston and LIGO-Hanford
spectrograms only in certain frequency ranges. The Virgo signal is not
visible. We indicated the expected position of the Virgo signal as a black
solid line in the Virgo spectrogram.

We constrain our analysis to GW frequencies below $150$ Hz. As mentioned
previously, in this range the orbital inspiral of solar mass neutron stars
is accurately described by the adiabatic approximation for which both vector
gravity and general relativity predict the same gravitational waveform $s(t)$%
. In particular, $s(t)$ can be approximated as 
\begin{equation}
s(t)\propto \frac{1}{\left( t_{c}-t\right) ^{1/4}}\cos \left[ 2\phi (t)%
\right] ,  \label{x1}
\end{equation}%
where 
\begin{equation}
\phi (t)=-\left( \frac{c^{3}}{5GM_{c}}\right) ^{5/8}\left( t_{c}-t\right)
^{5/8}+bt+\phi _{0},  \label{x2}
\end{equation}%
$t_{c}$ is the coalescence time and $M_{c}$ is the chirp mass of the system
given by Eq. (\ref{mc}). In Eq. (\ref{x2}) we introduced a small adjustable
frequency offset $b$ in order to obtain a better fit of the signal in a
broad frequency range. In Eqs. (\ref{x1}) and (\ref{x2}), $M_{c}$, $t_{c}$
and $b$ are free parameters chosen to give the maximum integrated signal (%
\ref{x3}). For the best fit we found in the detector frame $M_{c}=1.195$ $%
M_{\odot }$, $t_{c}=0.296$ s and $b=0.83$ rad/s\footnote{%
Use of a trial function with more variational parameters yields a better fit
of the phase of the detected GW. Adding the term $bt$, where $b$ is an
additional adjustable parameter, can increase the peak value of $I(t)$ by $%
20\%$ if we collect signal during $10$ s. Total change of the phase caused
by adding the term $bt$ is small. The point is that expression for $\phi (t)$
has other adjustable parameters ($M_{c}$ and $t_{c}$). Optimal values of
these parameters for $b=0.83$ rad/s are slightly different from that for $%
b=0 $. Their change compensates the main part of phase shift variation
produced by the term $bt$. The term $bt$, however, is not crucial. One can
do calculations without it and obtain similar results with slightly larger
uncertainty.}. The value of $M_{c}$ we obtained is close to that reported in
Refs. \cite{Abbo17b,Abbo18}.

\begin{figure}[h]
\centering
\hspace*{-0.25cm} \includegraphics[width=7.5cm,angle=270,origin=c]{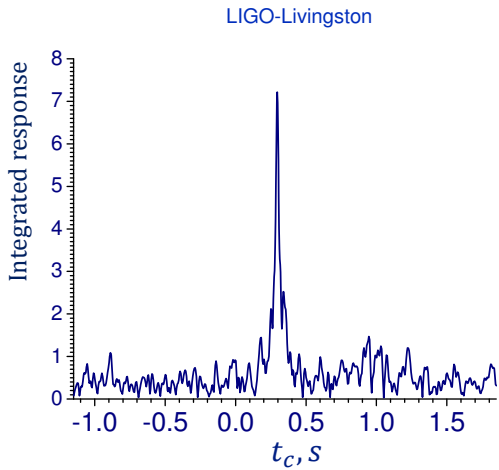}
\par
\vspace{-1.4cm}
\caption{Absolute value of the integrated interferometer response (\protect
\ref{x3}) (in arbitrary units) as a function of the coalescence time $t_{c}$
for the LIGO-Livingston detector accumulated from the frequency interval $%
51-115$ Hz. The response is calculated using the best fit signal phase $%
\protect\phi (t)$ given by Eq. (\protect\ref{x2}) with $M_{c}=1.195$ $%
M_{\odot }$ and $b=0.83$ rad/s. }
\label{Fig3}
\end{figure}

\begin{figure}[h]
\centering
\hspace*{-0.25cm} \includegraphics[width=7.5cm,angle=270,origin=c]{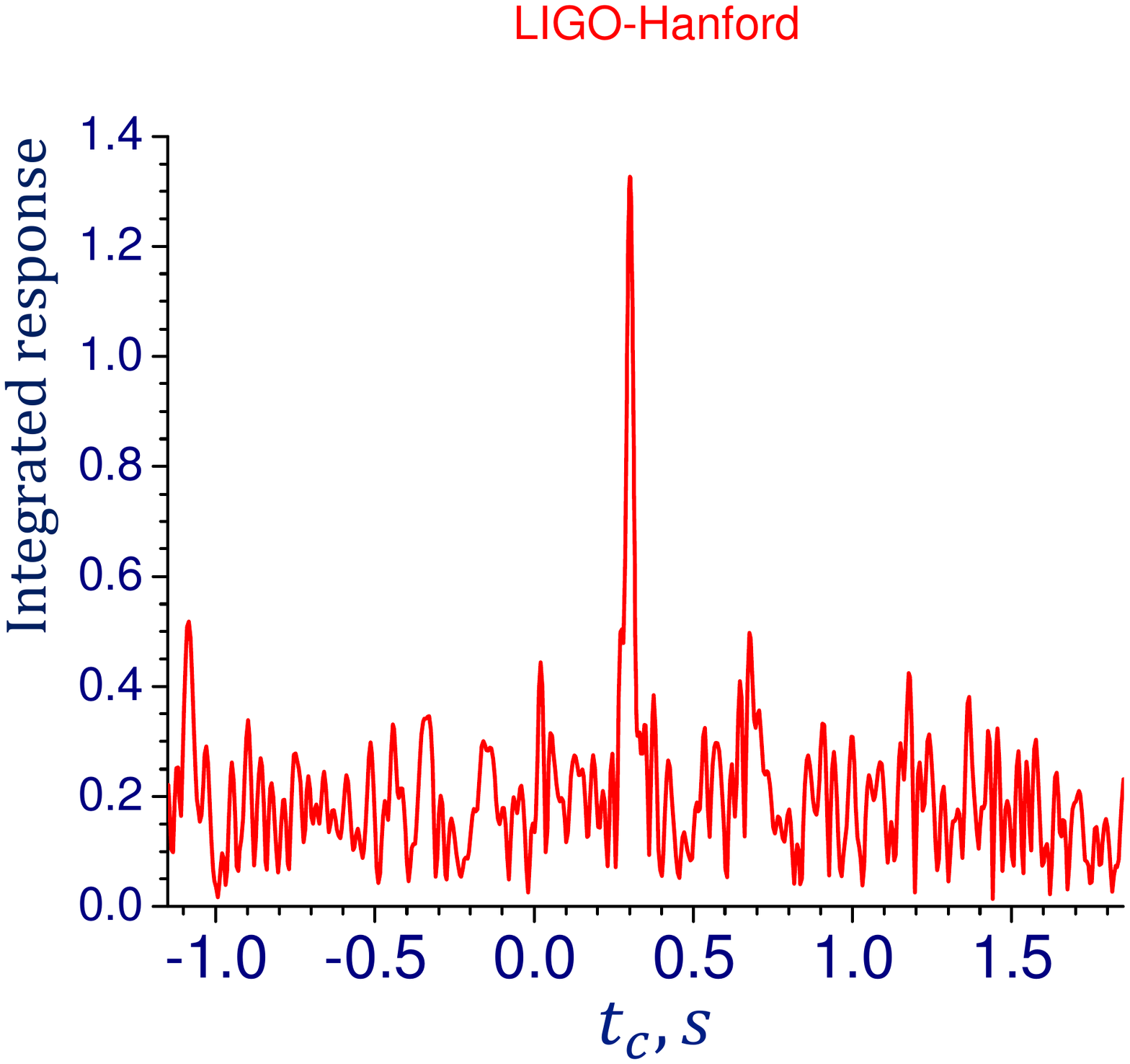}
\par
\vspace{-1.4cm}
\caption{Absolute value of the integrated interferometer response (\protect
\ref{x3}) (in arbitrary units) as a function of the coalescence time $t_{c}$
for the LIGO-Hanford detector accumulated from the frequency interval $%
85-145 $ Hz. The response is calculated using the best fit signal phase $%
\protect\phi (t)$ given by Eq. (\protect\ref{x2}) with $M_{c}=1.195$ $%
M_{\odot }$ and $b=0.83$ rad/s. }
\label{Fig4}
\end{figure}

\begin{figure}[h]
\centering
\hspace*{-0.25cm} \includegraphics[width=7.5cm,angle=270,origin=c]{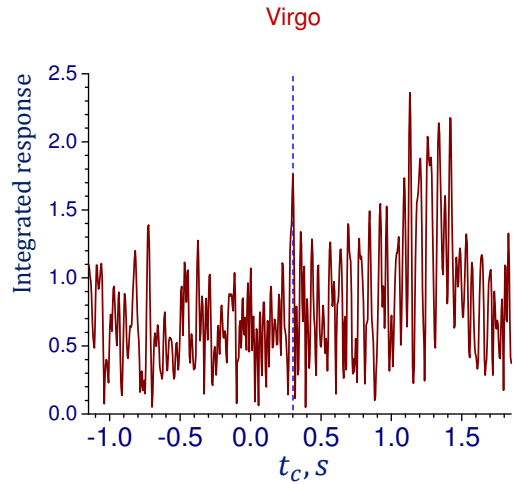}
\par
\vspace{-1.4cm}
\caption{Absolute value of the integrated interferometer response (\protect
\ref{x3}) (in arbitrary units) as a function of the coalescence time $t_{c}$
for the Virgo detector accumulated from the frequency interval $73-145$ Hz.
The response is calculated using the best fit signal phase $\protect\phi (t)$
given by Eq. (\protect\ref{x2}) with $M_{c}=1.195$ $M_{\odot }$ and $b=0.83$
rad/s. The vertical dashed line marks the expected position of the signal
peak.}
\label{Fig5}
\end{figure}

In Figs. \ref{Fig3}, \ref{Fig4} and \ref{Fig5} we plot the absolute value of
the integrated interferometer response $|I|$ (given by Eq. (\ref{x3})) as a
function of the coalescence time $t_{c}$ for the three interferometers. The
plots have a pronounced peak when $t_{c}=0.296$ s for LIGO-Livingston and
LIGO-Hanford detectors. Thus, for these detectors the signal can be found
with a high accuracy. For the Virgo detector the integrated signal is barely
visible on top of the noise background (see Fig. \ref{Fig5}).

\begin{figure}[h]
\centering
\hspace*{-0.25cm} \includegraphics[width=7.5cm,angle=270,origin=c]{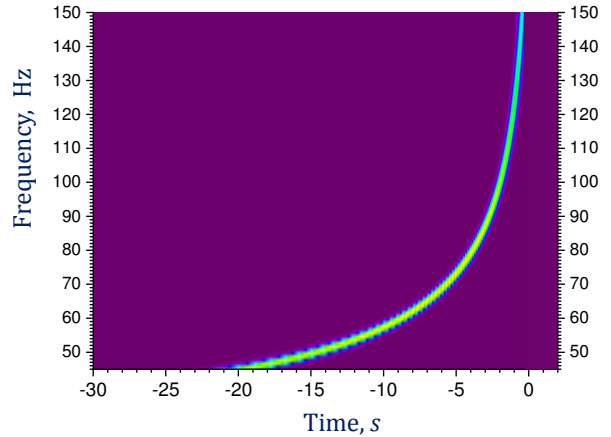}
\par
\vspace{-1.4cm}
\caption{Spectrogram of the best fit gravitational waveform given by Eqs. (%
\protect\ref{x1}) and (\protect\ref{x2}) with $M_{c}=1.195$ $M_{\odot }$, $%
t_{c}=0.296$ s and $b=0.83$ rad/s. Color coding is the same as in Fig. 
\protect\ref{Fig2}.}
\label{Fig6}
\end{figure}

\begin{figure}[h]
\centering
\hspace*{-0.5cm} \includegraphics[width=7.5cm,angle=270,origin=c]{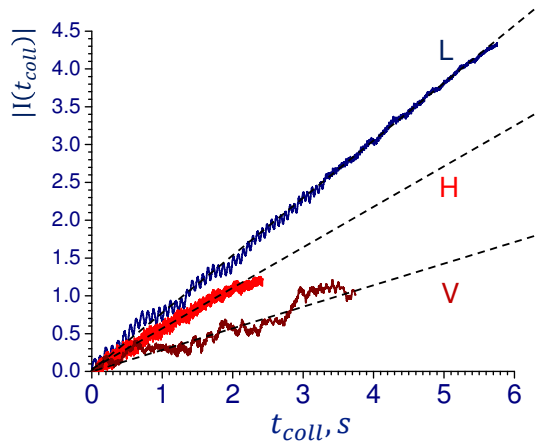}
\par
\vspace{-1.4cm}
\caption{Absolute value of the integrated signal (\protect\ref{x3}) as a
function of the collection time $t_{\text{coll}}$ for LIGO-Livingston (upper
curve), LIGO-Hanford (middle curve) and Virgo (bottom curve) detectors. The
scale of the vertical axis is arbitrary but the same for all three
detectors. The signals are collected from the frequency intervals (\protect
\ref{x4}), (\protect\ref{x5}) and (\protect\ref{x6}) respectively. The
least-squares fits to the data are shown as dashed lines. }
\label{Fig7}
\end{figure}

In Fig. \ref{Fig6} we plot a spectrogram of the best fit gravitational
waveform given by Eqs. (\ref{x1}) and (\ref{x2}). In the time-frequency
representation such an idealized signal is a smooth line. In contrast, the
LIGO-Livingston signal contains gaps in the spectrogram at certain
frequencies (see Fig. \ref{Fig2} (top) and Fig. 1 in Ref. \cite{Abbo17b}).
Presumably, these \textquotedblleft gaps\textquotedblright\ are due to
real-time noise filtering through various feedback loops and off-line noise
subtraction applied to the LIGO-Livingston detector \cite{Abbo17b,Abbo18}.
Such a procedure eliminates noise at certain frequencies and, as one can see
from the spectrogram, it also reduces the LIGO-Livingston signal at these
frequencies. As a consequence, only frequency ranges for which the signal is
clearly visible in the spectrogram should be used for the estimate of the
LIGO-Livingston signal amplitude (for details see Section \ref{Comment}). We
emphasize that the $H/L$ and $V/L$ ratios should be the same throughout the
early inspiral stage. We accumulate the LIGO-Livingston signal from the
following \textquotedblleft good\textquotedblright\ frequency intervals 
\begin{equation}
L:\quad 51\div 57,\quad 71\div 80,\quad 95\div 115\,\text{ Hz.}  \label{x4}
\end{equation}%
This means that when we calculate the integrated interferometer response (%
\ref{x3}) we integrate only over the time intervals for which the signal
passes through the frequency bands (\ref{x4}). This yields a total
collection time of about $6$ s. The result for $|I(t)|$ as a function of the
collection time $t_{\text{coll}}$ is shown in Fig. \ref{Fig7} (upper curve).
The figure shows that the absolute value of the integrated LIGO-Livingston
signal $|I(t)|$ approximately follows a straight line, in agreement with
Eqs. (\ref{f1a}) and (\ref{f2a}).

\begin{figure}[h]
\centering
\hspace*{-0.4cm} \includegraphics[width=7.5cm,angle=270,origin=c]{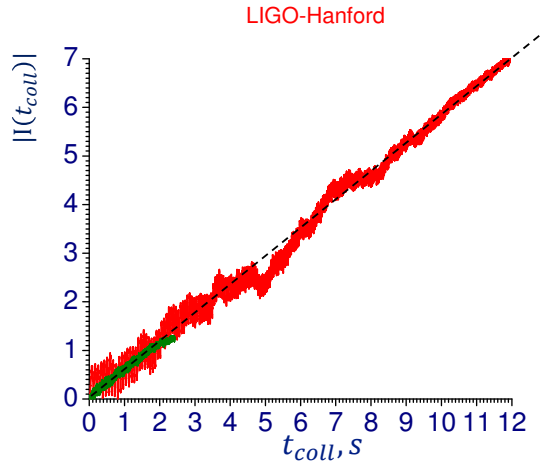}
\par
\vspace{-1.4cm}
\caption{Absolute value of the integrated signal (\protect\ref{x3}) as a
function of the collection time $t_{\text{coll}}$ for the LIGO-Hanford
detector collected from the frequency intervals (\protect\ref{x5}) (green
curve) and $51\div 115\,$ Hz (red curve). The fit to the data is shown as a
dashed line. }
\label{Fig8}
\end{figure}

As we show later in Section \ref{Comment}, the noise filtering has not
appreciably affected the LIGO-Hanford signal and hence, in principle, one
can use a broad frequency range for the LIGO-Hanford signal collection.
However, in our estimate we collect the signal only from frequency intervals
with lower noise. Namely, we choose 
\begin{equation}
H:\quad 85\div 121,\quad 128\div 145\,\text{ Hz.}  \label{x5}
\end{equation}%
This yields a collection time of about $2.5$ s. The result for the
LIGO-Hanford integrated signal $|I(t)|$ is shown as the middle curve in Fig. %
\ref{Fig7}, which also follows a straight line. As a check, in Fig. \ref%
{Fig8} we show that the LIGO-Hanford signal collection from a broad
frequency range of $51\div 115\,$ Hz (collection time $12$ s) yields the
same average slope of $|I(t)|$. Thus, the Hanford results obtained using
different frequency intervals are consistent with each other.

Due to the larger amplitude noise in the Virgo observatory, the GW signal is
not visible in the Virgo spectrogram. To estimate the Virgo signal we
integrate the measured strain over the frequency ranges corresponding to the
lowest detector noise at the expected time of the signal arrival. Namely, we
accumulate signal from the following frequency intervals 
\begin{equation}
V:\quad 73\div 85,\quad 89\div 122,\quad 130\div 145\,\text{ Hz}  \label{x6}
\end{equation}%
The result is shown as the bottom curve in Fig. \ref{Fig7}. The curve does
not follow a straight line and, hence, we cannot estimate the Virgo signal
with a good accuracy. However, the plot allows us to place a reliable upper
limit on the Virgo signal amplitude.

\begin{figure}[h]
\centering
\hspace*{-0.25cm} \includegraphics[width=7.5cm,angle=270,origin=c]{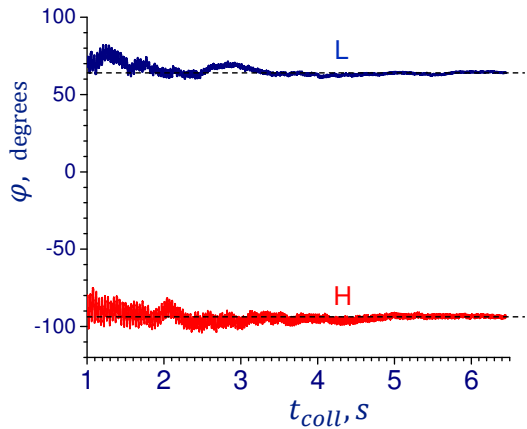}
\par
\vspace{-1.4cm}
\caption{Phase of the integrated signal (\protect\ref{x3}) as a function of
the collection time for the LIGO-Livingston (upper curve) and LIGO-Hanford
(lower curve) strain data. The signals are collected from the frequency
range of $62\div 115\,$ Hz for both detectors. Dashed lines show the best
fit phase values. Since we adjusted the measured strain time series for the
arrival time delays, they do not contribute to the phase differences.}
\label{Fig9}
\end{figure}

In Fig. \ref{Fig9} we plot the phase of the integrated signal $I(t)$ as a
function of the collection time for the LIGO-Livingston and LIGO-Hanford
strain data. Contrary to the amplitude, the phase of the accumulated signal
is not affected if the signal is reduced at certain frequencies by the noise
filtering. For the phase estimate we collected the interferometer signals in
the same frequency range of $62\div 115\,$ Hz (total collection time is $6.5$
s) for both detectors. As expected, the accumulated phase is approximately
constant as a function of the collection time and can be obtained with a
good accuracy from the plots of Fig. \ref{Fig9}.

Figures \ref{Fig7} and \ref{Fig9} yield the following estimates%
\begin{equation}
\left\vert \frac{H}{L}\right\vert =0.65\pm 0.15,  \label{f8}
\end{equation}%
\begin{equation}
\varphi _{L}-\varphi _{H}=160^{\circ }\pm 15^{\circ },  \label{f8a}
\end{equation}%
\begin{equation}
\left\vert \frac{V}{L}\right\vert <0.45,  \label{f9}
\end{equation}%
where the uncertainties correspond to $2\sigma $ confidence interval ($95\%$
confidence level). The uncertainties have been calculated by injecting a
test signal into the measured strain time series.

According to Eqs. (\ref{f1}) and (\ref{f2}), one can check the consistency
of our estimates by calculating the ratio of the Fourier transforms of the
signals in different detectors. In order to obtain the signal in the
time-frequency representation we calculate the short-time Fourier transform
of the measured time series $h(t)$. To do so, we divide time into intervals
of length $\Delta t=0.3$ s. This is an optimal value of $\Delta t$; it
covers a large enough number of GW oscillations and yet it is short enough
not to wash out the signal. We calculate the Fourier transform for each time
interval $[t,t+\Delta t]$ 
\begin{equation}
F(f,t)=\sum\limits_{t<t_{k}<t+\Delta t}h(t_{k})e^{-2\pi ift_{k}}
\end{equation}%
and plot $F(f,t)$ at fixed frequency $f$ as a function of time by changing $%
t $ in steps much shorter than $\Delta t$.

\begin{figure}[h]
\centering
\hspace*{-0.25cm} \includegraphics[width=7.5cm,angle=270,origin=c]{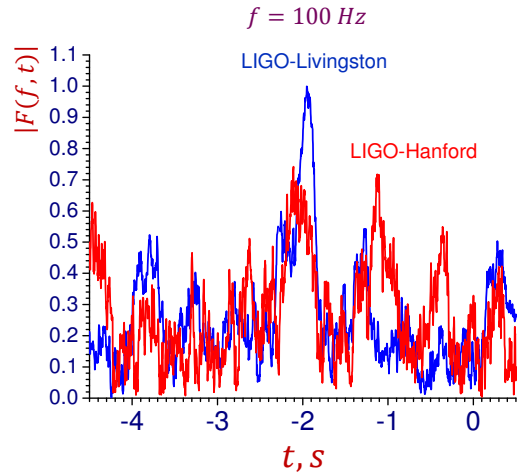}
\par
\vspace{-1.5cm}
\caption{Absolute value of the Fourier transform of the strain $|F(f,t)|$
for the GW170817 event measured by the LIGO-Hanford (red line) and
LIGO-Livingston (blue line) interferometers as a function of time at
frequency $f=100$ Hz. The vertical axis is normalized such that the
LIGO-Livingston GW signal amplitude is equal to $1$.}
\label{Fig10}
\end{figure}

\begin{figure}[h]
\centering
\hspace*{-0.25cm} \includegraphics[width=7.5cm,angle=270,origin=c]{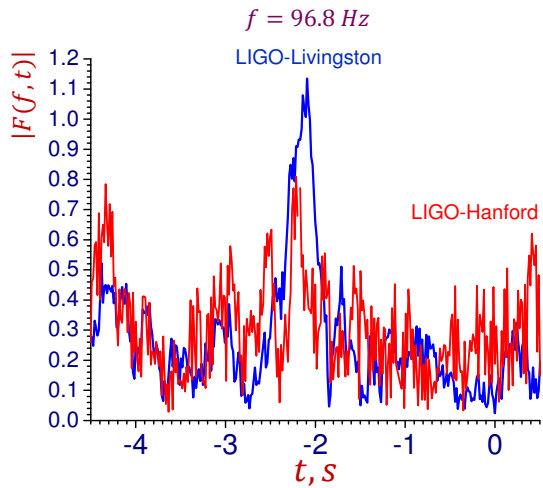}
\par
\vspace{-1.4cm}
\caption{Absolute value of the Fourier transform of the strain $|F(f,t)|$
for the GW170817 event measured by the LIGO-Hanford (red line) and
LIGO-Livingston (blue line) interferometers as a function of time at
frequency $f=96.8$ Hz. The scale of the vertical axis is the same as in Fig. 
\protect\ref{Fig10}.}
\label{Fig11}
\end{figure}

\begin{figure}[h]
\centering
\hspace*{-1cm} \includegraphics[width=8cm,angle=270,origin=c]{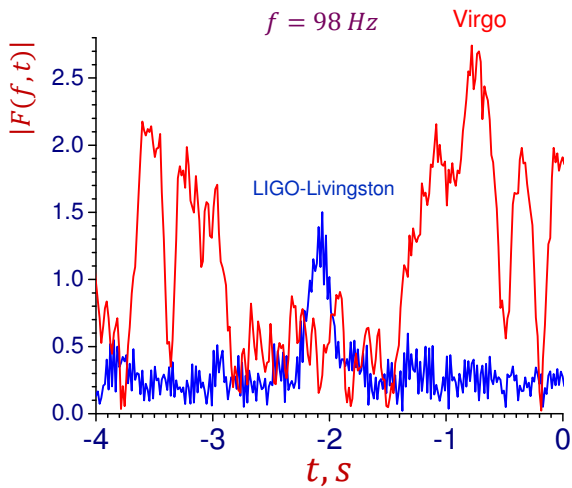}
\par
\vspace{-1.4cm}
\caption{ Absolute value of the Fourier transform of the strain $|F(f,t)|$
measured by the Virgo (red line) and LIGO-Livingston (blue line)
interferometers as a function of time at frequency $f=98$ Hz. The scale of
the vertical axis is the same as in Fig. \protect\ref{Fig10}.}
\label{Fig12}
\end{figure}

The GW signal is clearly visible in the LIGO-Livingston data only over a few
frequency ranges. Among them we select frequencies for which the noise of
the LIGO-Hanford detector is relatively small. In Fig. \ref{Fig10} we plot
the absolute value of the Fourier transform of the strain $|F(f,t)|$
measured by the LIGO-Livingston (blue curve) and the LIGO-Hanford (red
curve) interferometers as a function of time at frequency $f=100$ Hz. The GW
signal is clearly visible at $t_{0}\approx -2$ s. The vertical axis in the
figure is normalized such that the LIGO-Livingston signal amplitude is equal
to $1$.

Fig. \ref{Fig10} shows that $|H/L|$ is consistent with the estimate (\ref{f8}%
). We found that similar plots for other frequencies at which the
LIGO-Livingston signal is clearly visible and the LIGO-Hanford noise level
is low give the same answer. For example, Fig. \ref{Fig11} shows the
absolute value of the Fourier transform of the strain $|F(f,t)|$ measured by
the LIGO-Hanford and LIGO-Livingston interferometers as a function of time
at frequency $f=96.8$ Hz. The plots shown in Fig. \ref{Fig11} are consistent
with Eq. (\ref{f8}).

For the Virgo detector the noise is very high, which makes extraction of the
Virgo signal a challenging task. Using Virgo data we calculated $F(f,t)$ at
various frequencies $f$ and luckily found one which can be used for the
signal estimate. We found that for $f=98$ Hz the Virgo noise happened to be
quite low in the vicinity of the time at which the signal is expected to
arrive (see Fig \ref{Fig12}). As a result, using $F(f,t)$ for $f=98$ Hz we
can constrain the $V/L$ ratio with a reasonable accuracy which yields an
estimate consistent with Eq. (\ref{f9}).

\section{Test of gravitational theories}

\label{Test}

According to Eqs. (\ref{f10}) and (\ref{f11}), the complex ratios of the GW
signals detected by the different interferometers depend on whether the GW
is described by vector gravity or general relativity. The right hand sides
of Eqs. (\ref{f10}) and (\ref{f11}) contain only two unknown parameters, the
inclination and polarization angles $\theta $ and $\psi $.

According to the results of the previous section, the experimental
constraints on the complex ratios $H/L$ and $V/L$ are given by Eqs. (\ref{f8}%
), (\ref{f8a}) and (\ref{f9}) respectively. If there are angles $\theta $
and $\psi $ for which Eq. (\ref{f10}) and the corresponding equation for $%
V/L $ yield the measured ratios, then vector gravity agrees with the
observations. Otherwise, the theory is ruled out. Eq. (\ref{f11}) and the
corresponding equation for $V/L$ can be used to test general relativity. As
mentioned previously, only one of the two theories is expected to pass this
test.

\subsection{Test of vector gravity}

\begin{figure}[h]
\centering
\hspace*{-0.5cm} \includegraphics[width=7.5cm,angle=270,origin=c]{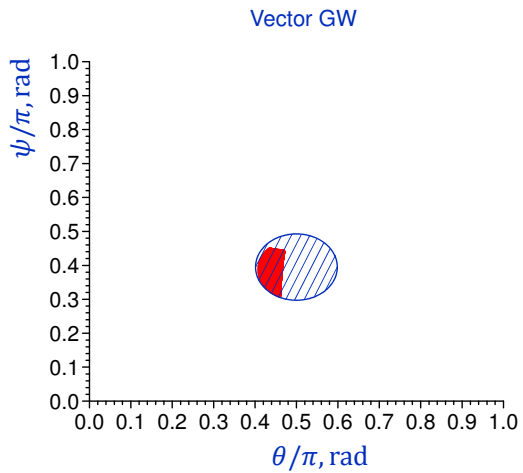}
\par
\vspace{-1.4cm}
\caption{The region of inclination $\protect\theta $ and polarization $%
\protect\psi $ angles compatible with constraints (\protect\ref{f8}), (%
\protect\ref{f8a}) and (\protect\ref{f9}) for the GW170817 event assuming
vector GW (red filled area). The dashed area is the region of angles
calculated in the vector theory of gravity which gives a distance to the
source compatible with the astronomical observations. The allowed
inclination angle range is $71^{\circ }<\protect\theta <84^{\circ }$.}
\label{Fig13}
\end{figure}

We found that for vector gravity there is a range of inclination $\theta $
and polarization $\psi $ angles compatible with the constraints (\ref{f8}), (%
\ref{f8a}) and (\ref{f9}). This range is shown in Fig. \ref{Fig13} as the
red filled area. Thus, vector gravity is compatible with the observations.

Next we show that vector gravity also gives the correct distance to the
source. Eq. (\ref{s1}) tells us that in vector gravity the intensity of GWs
emitted at the inclination angle $\theta $ is proportional to%
\begin{equation}
I_{VG}\propto \sin ^{2}\theta \left( 1+\cos ^{2}\theta \right) ,
\end{equation}%
that is, the GW emission is maximum in the orbital plane ($\theta =\pi /2$)
and is equal to zero perpendicular to the plane ($\theta =0,$ $\pi $). On
the other hand, for general relativity we have \cite{Land95}%
\begin{equation}
I_{GR}\propto 4\cos ^{2}\theta +(1+\cos ^{2}\theta )^{2}.
\end{equation}%
The emission intensity peaks in the direction perpendicular to the orbital
plane ($\theta =0,$ $\pi $) and drops by a factor of eight for the in-plane
emission.

NGC 4993 located at a distance $43.8_{-6.9}^{+2.9}$ Mpc was identified as
the host galaxy of GW170817 \cite{Abbo17b}. It has been shown that a general
relativistic GW yields the right distance to the source at the $95.4\%$ ($%
2\sigma $) credible level if $|\cos \theta |>0.75$ \cite{Abbo17c}. Using
Eqs. (\ref{f1}) and (\ref{f2}) we calculated the region of $\theta $ and $%
\psi $ angles for which the amplitude of the $L$ signal in vector gravity is
equal to that produced by a general relativistic GW coming from the same
binary system with $|\cos \theta |>0.75$. Direct calculations based on Eqs. (%
\ref{s1}) and (\ref{aa}), known parameters of the binary system \cite{Abbo18}
and the amplitude of the $L$ signal yield similar results.

The constraints on $\theta $ and $\psi $ obtained in vector gravity based on
the known distance to the source are shown in Fig. \ref{Fig13} as the dashed
area. The dashed and filled red regions have considerable overlap. Thus,
vector gravity yields a distance to the source compatible with the
astronomical observations, but with a different conclusion about the binary
orbit's inclination angle.

One should note that Fig. \ref{Fig13} constrains the orbit inclination angle 
$\theta $ to the range $71^{\circ }<\theta <84^{\circ }$, that is, according
to vector gravity, the line-of-sight is close to the orbital plane of the
inspiraling stars. This suggests that the faint short gamma-ray burst
detected $1.7$ s after GW170817 was not produced by a canonical relativistic
jet viewed at a small angle. The canonical jet scenario is also in tension
with the peak energy of the observed gamma-ray spectrum \cite{Ioka19}. The
detected gamma-ray burst probably was produced by a different mechanism,
e.g. by elastic scattering of the jet radiation by a cocoon \cite%
{Kisa15,Kisa17,Kisa18} or by a shock breakout of the cocoon from the
merger's ejecta \cite{Kasl17,Gott18,Naka18,Ioka19a}.

Later observations with VLBI discovered a compact radio source associated
with the GW170817 remnant exhibiting superluminal apparent motion between
two epochs at $75$ and $230$ days post-merger \cite{Mool18}. This indicates
the presence of energetic and narrowly collimated ejecta observed from an
angle of about $20$ degrees at the late-time emission \cite{Mool18}.
Typically ejecta of spinning neutron stars consist of polar jets directed
along the pulsar spin-axis and a much brighter equatorial wind. Since vector
gravity constrains the line-of-sight close to the orbital plane the observed
compact superluminal source is probably associated with the stellar
equatorial wind.

\subsection{Test of general relativity}

\begin{figure}[h]
\centering
\hspace*{-0.5cm} \includegraphics[width=8cm,angle=270,origin=c]{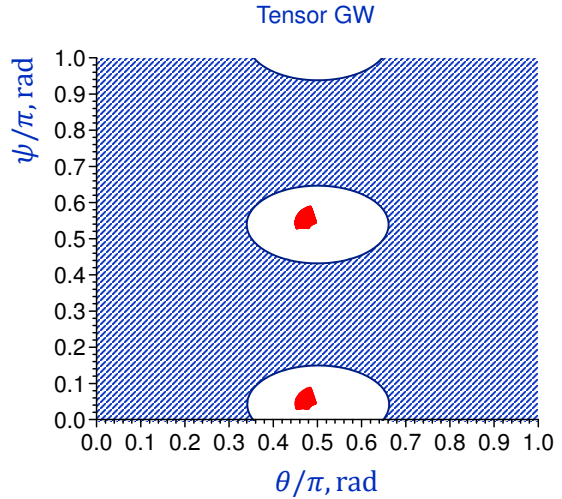}
\par
\vspace{-1.5cm}
\caption{The range of inclination $\protect\theta $ and polarization $%
\protect\psi $ angles for general relativistic GWs compatible with
constraints (\protect\ref{f8}), (\protect\ref{f8a}) (red filled region) and (%
\protect\ref{f9}) (blue shaded region). The red and blue regions do not
overlap. Thus, there is no range of $\protect\theta $ and $\protect\psi $
compatible with constraints (\protect\ref{f8}), (\protect\ref{f8a}) and (%
\protect\ref{f9}) simultaneously.}
\label{Fig14}
\end{figure}

We found that for a general relativistic tensor GW there are no combinations
of inclination $\theta $ and polarization $\psi $ angles compatible with
constraints (\ref{f8}), (\ref{f8a}) and (\ref{f9}) (see Fig. \ref{Fig14}).
Uncertainties in Eqs. (\ref{f8}), (\ref{f8a}) and (\ref{f9}) correspond to $%
2\sigma $ confidence interval. We found that general relativity is
incompatible with constraints (\ref{f8}), (\ref{f8a}) and (\ref{f9}) if the
uncertainties are increased upto $2.5\sigma $. Therefore, the general theory
of relativity is inconsistent with the data at the $2.5\sigma $ ($99\%$
confidence) level.

\begin{figure}[h]
\centering
\hspace*{-1cm} \includegraphics[width=8cm,angle=270,origin=c]{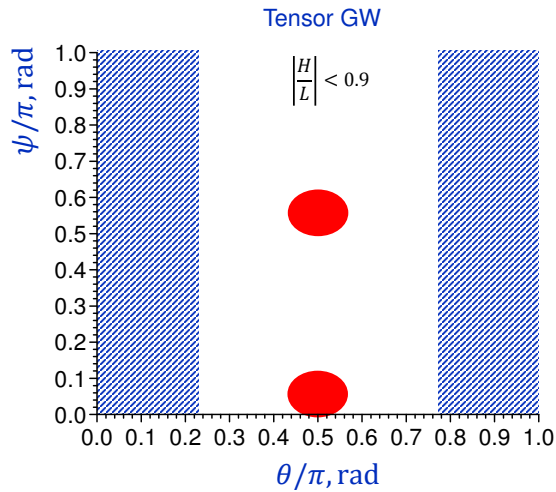}
\par
\vspace{-1.4cm}
\caption{The region of inclination $\protect\theta $ and polarization $%
\protect\psi $ angles compatible with the single constraint $|H/L|<0.9$ for
the GW170817 event assuming general relativistic GWs (red filled ovals). No
constraint on $V/L$ was imposed. The blue shaded rectangles show the region
of inclination angles which give a distance to the source compatible with
the astronomical observations.}
\label{Fig15}
\end{figure}

A skeptical reader might argue that constraints (\ref{f8}), (\ref{f8a}) and (%
\ref{f9}) are too restrictive or the procedure we used is not accurate.
Anticipating such criticism, we next show that general relativity is at odds
with observations even if we estimate the $|H/L|$ ratio directly from the
Fourier transform of the gravitational-wave strain data without using the
signal accumulation algorithm. Plots of the Fourier transform in Figs. \ref%
{Fig10} and \ref{Fig11} show that at least $|H/L|<0.9$. Note that the
short-time Fourier transform method complements the signal accumulation
method because it is independent of the details of the waveform model.

It turns out that general relativity can be ruled out even based on the
single constraint $|H/L|<0.9$. In Fig. \ref{Fig15} we plot the region of
inclination $\theta $ and polarization $\psi $ angles compatible with the
GW170817 event assuming general relativistic GWs and $|H/L|<0.9$ (red filled
ovals). No constraint on $V/L$ was imposed. The blue shaded rectangle
regions indicate inclination angles which yield a distance to the source
compatible with the astronomical observations at the $2\sigma $ level ($%
|\cos \theta |>0.75$ \cite{Abbo17c}). Since the blue and red regions do not
overlap, we conclude that general relativistic GW can not simultaneously
give the right distance to the source and be compatible with the measured
ratio $|H/L|$.

The constraint on the ratio $|H/L|$ is crucial for distinguishing between
general relativity and vector gravity. Usually the importance of this
constraint is not appreciated due to common belief that the two LIGO
instruments are nearly co-aligned and, thus, $|H/L|$ can not give additional
constraint on the GW polarization \cite{Isi17}. As we mentioned in the
Introduction, the angles between the corresponding arms of the two LIGO
interferometers are actually not that small ($13^{\circ }$ and $24^{\circ }$
respectively) and hence the $H/L$ ratio can impose significant constraints.

\begin{figure}[h]
\centering
\hspace*{-0.5cm} \includegraphics[width=7.5cm,angle=270,origin=c]{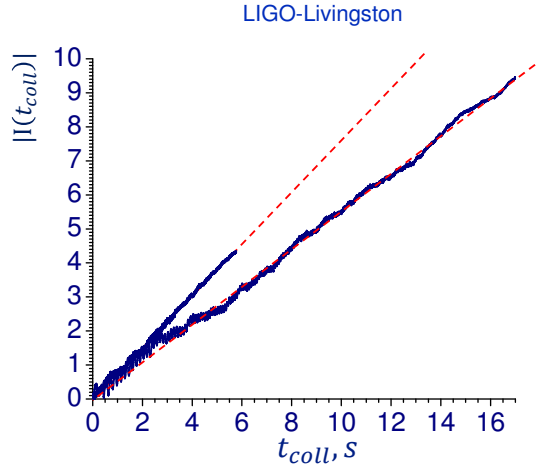}
\par
\vspace{-1.4cm}
\caption{Absolute value of the integrated signal (\protect\ref{x3}) as a
function of the collection time $t_{\text{coll}}$ for the LIGO-Livingston
detector collected from the \textquotedblleft good\textquotedblright\
frequency intervals (\protect\ref{x4}) (upper curve) and from a wide
frequency range $45\div 115\,$ Hz (lower curve). The best fit curves are
shown as dashed lines.}
\label{Fig16}
\end{figure}

We want to emphasize that in order to avoid a systematic error in
determining the LIGO-Livingston signal amplitude one should accumulate the
signal only from the frequency intervals for which the signal is clearly
visible in the spectrogram. This eliminates contributions from the regions
at which signal was reduced by noise filtering. Not doing so can result in a
substantial underestimate of the LIGO-Livingston signal amplitude as
demonstrated in Fig. \ref{Fig16}. The figure shows that the integrated
signal $|I(t)|$ collected from a wide frequency range $45\div 115$ Hz does
not follow a straight line and yields substantially smaller average slope of
the $|I(t)|$ curve (the average signal per unit collection time).

An underestimate of the LIGO-Livingston signal increases the $|H/L|$ ratio
which can mimic agreement with general relativity. We believe that such an
underestimation error has been made in the LIGO-Virgo polarization analysis
of GW170817 \cite{Max18} which we discuss next.

\section{Comment on LIGO-Virgo polarization analysis of GW170817}

\label{Comment}

In a recent paper \cite{Max18} the LIGO-Virgo collaboration reported results
of the GW polarization test with GW170817 performed using a Bayesian
analysis of the signal properties with the three LIGO-Virgo interferometer
outputs. The authors found overwhelming evidence in favor of pure tensor
polarization over pure vector with an exponentially large Bayes factor. This
result is opposite to our present findings. Here we explain why we came to
the opposite conclusion and argue that the results reported in \cite{Max18}
should be reconsidered by removing the depleted\ frequency intervals from
the LIGO-Livingston strain time series.

According to Eqs. (\ref{f1a}) and (\ref{f2a}) the integrated interferometer
response $I(t)$ grows linearly with the collection time $t$ and the phase of 
$I(t)$ is independent of $t$. Thus, the theory predicts that the ratio 
\begin{equation}
u=\frac{I(t)}{t}
\end{equation}%
should be independent of the signal collection time interval $%
[t_{0},t_{0}+t] $. This ratio can be interpreted as a signal per unit time. $%
u$ can be used to determine how much signal is present in the interferometer
data stream at different times. If noise filtering has not altered the
signal at certain frequencies then $u$ should have the same (complex) value
for any collection time interval.

\begin{figure}[h]
\centering
\hspace*{-0.5cm} \includegraphics[width=7.5cm,angle=270,origin=c]{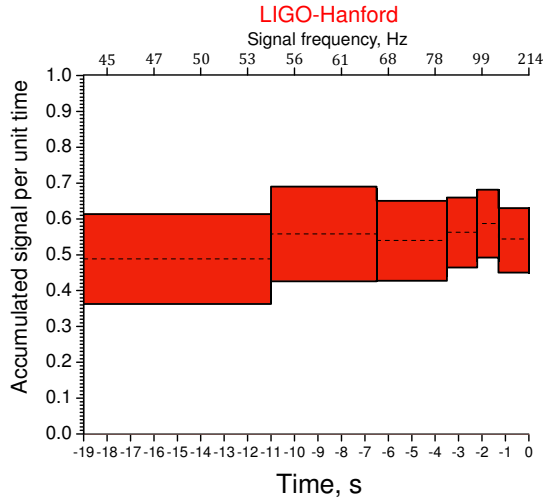}
\par
\vspace{-1.4cm}
\caption{Absolute value of the signal per unit time $|u|$ measured by
LIGO-Hanford detector for the GW170817 event for different collection time
intervals. As in Ref. \protect\cite{Abbo17b}, times are shown relative to
August 17, 2017 12:41:04 UTC.}
\label{CFig1}
\end{figure}

\begin{figure}[h]
\centering
\hspace*{-0.2cm} \includegraphics[width=7.5cm,angle=270,origin=c]{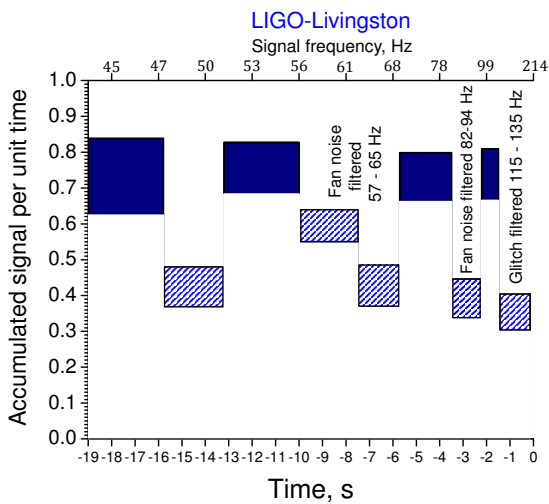}
\par
\vspace{-1.4cm}
\caption{ The same as in Fig. \protect\ref{CFig1}, but for the
LIGO-Livingston detector. The scale of the vertical axis is arbitrary but
the same in both figures. Frequency regions in which the signal is reduced
by noise filtering are indicated as dashed bars.}
\label{CFig2}
\end{figure}

In Fig. \ref{CFig1} we plot $|u|$ for LIGO-Hanford detector for different
collection time intervals. The result is shown as a set of rectangular bars.
The length of a bar is equal to the collection time $t$, while the height
corresponds to the uncertainty produced by the detector noise. Namely, the
half-height is equal to one standard deviation. We calculated the
uncertainty by injecting a test signal into the measured strain time series
before and after the GW170817 event. The average value of $|u|$ for each
interval $t$ is indicated by a dashed line. The figure shows that $|u|$ for
the LIGO-Hanford detector is consistent with a constant for the entire time
when the signal is present in the data stream. Therefore, the signal in the
LIGO-Hanford strain time series appears not to be affected by the noise
mitigation.

In Fig. \ref{CFig2} we plot $|u|$ for LIGO-Livingston detector. The
uncertainty bars for this detector are somewhat smaller due to lower noise.
Solid color bars indicate regions for which signal is clearly visible in the
LIGO-Livingston spectrogram (see Fig. \ref{Fig2} top panel). The vertical
position of the solid color bars is consistent with $u$ being a constant, as
predicted by the theory. Dashed bars correspond to gap regions in the
spectrogram. Fig. \ref{CFig2} shows that the signal in these regions is
substantially smaller than the signal content in other intervals.

The signal reduction can be attributed to noise removal which is partially
performed in real time using various feedback loops. \textquotedblleft
Before noise subtraction\textquotedblright\ strain data published by the
LIGO-Virgo collaboration \cite{LV17} already went through the real-time
noise removal which mainly involves filtering of mechanical noise. In the
\textquotedblleft after noise subtraction\textquotedblright\ data some
additional noise contributions have been eliminated. Namely, the off-line
noise subtraction removed 60 Hz AC power mains harmonics from the
LIGO-Livingston data stream and a glitch which occurred in the detector
about $1.1$ s before coalescence. The latter can explain why there is less
signal in the data stream during the glitch duration. Namely, the off-line
glitch removal from the data stream led to a reduction of the GW signal in
that time segment, as one can see from Fig. \ref{CFig2}. The signal
reduction in other frequency regions of the LIGO-Livingston detector can be
attributed to real-time filtering of fan noise (see Fig. 3 in \cite{Cost18}).

Usually it is believed that noise filtering does not reduce the signal
substantially. But this is just an assumption which must be tested in an
experiment with real gravitational waves. Because GW170817 involves orbital
inspiral of low-mass stars the GW signal slowly passes through the detector
frequency band. This allows us to determine the amount by which GW signal is
suppressed by the noise removal at different frequencies. It seems possible
that the noise filtering yields substantial signal reduction at certain
frequencies for GW events for which the GW signal per unit time is very
weak, which is the case for GW170817. Thus, for such events one should
perform a consistency check of Figs. \ref{CFig1} and \ref{CFig2}, and remove
\textquotedblleft corrupted\textquotedblright\ frequency intervals from the
data analysis. This must be taken into account in the analysis of future GW
events produced by inspiral of low-mass objects.

Contrary to the amplitude situation, we found that the signal phase was not
altered in the Livingston and Hanford detectors, namely the phase of $u$ is
consistent with a constant (see Fig. \ref{Fig9}). Source triangulation maybe
obtained using differences in arrival times of the signal in various
detectors \cite{Fair18}. For the GW170817 event the Hanford-Livingston delay
can be found with a high accuracy by fitting the signal phase $\phi (t)$ for
hundreds of GW cycles. Since the signal phase was not altered by noise
filtering the source triangulation from the Hanford-Livingston delay is
predicted correctly.

As we showed in the previous section, the ratio of LIGO-Hanford and
LIGO-Livingston signal amplitudes $|H/L|$ is crucial for distinguishing
between tensor and vector GW polarizations. In Section \ref{Data} we
obtained this ratio by accumulating the LIGO-Livingston signal only from the
regions shown by the solid color bars in Fig. \ref{CFig2}. In these regions
the signal is not depleted. We found $0.5<\left\vert H/L\right\vert <0.8$ at
the $2\sigma $ confidence level. This estimate, combined with constraints (%
\ref{f8a}) and (\ref{f9}), is compatible with the vector theory of gravity 
\cite{Svid17,Svid18} but rules out general relativity.

\begin{figure}[h]
\centering
\hspace*{-0.5cm} \includegraphics[width=8cm,angle=270,origin=c]{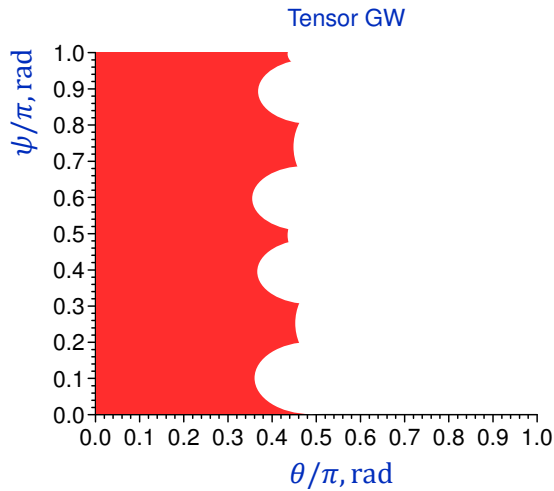}
\par
\vspace{-1.4cm}
\caption{Range of inclination $\protect\theta $ and polarization $\protect%
\psi $ angles for general relativistic GWs compatible with constraints $%
1.1<|H/L|<1.3$, $\protect\varphi _{L}-\protect\varphi _{H}=160^{\circ }\pm
15^{\circ }$ and $|V/L|<0.77$ for the GW170817 event (red filled region).}
\label{Fig17}
\end{figure}

However, signal accumulation from the entire frequency band erroneously
underestimates the LIGO-Livingston signal amplitude yielding $1.1<\left\vert
H/L\right\vert <1.3$ and $\left\vert V/L\right\vert <0.77$ \cite{Remark}.
This incorrect estimate results in the opposite conclusion about GW
polarization. Namely, it rules out vector polarization, but is consistent
with tensor polarization (see Fig. \ref{Fig17}). This is what the authors of
Ref. \cite{Max18} have found. The incorrect estimate of $|L|$ also explains
why the sky location of the source is nevertheless predicted correctly with
tensor polarization (general relativity).

One should mention that inconsistencies in the LIGO-Livingston data at
different GW frequencies have been also found in the Bayesian estimation of
the binary tidal deformability $\tilde{\Lambda}$ \cite{Nari18}. Namely it
has been shown that the probability distribution for $\tilde{\Lambda}$
obtained from the LIGO-Livingston strain time series changes irregularly
under variation of the maximum frequency of the data used in the analysis.
These inconsistent features are not observed for the Hanford detector \cite%
{Nari18}.

We also performed polarization analysis in a Bayesian framework (that has
been used by the LIGO-Virgo collaboration) taking into account the
Livingston signal amplitude reduction in certain frequency regions. We found
that the data give a $10^{8}$ Bayes factor favoring vector polarization over
tensor.

The same method shows that if the analysis uses the smaller Livingston GW
amplitude that results from including the signal depleted regions, then the
Bayes factor favors tensor polarization by an exponentially large factor.
That is what the LIGO-Virgo collaboration claims. The Bayesian analysis
confirms that getting the relative amplitude ratios correct is absolutely
critical and that if we use the \textquotedblleft correct\textquotedblright\
amplitude ratios, the GW170817 data strongly favor vector polarization.

\section{Null stream analysis}

Our conclusion about vector polarization of GWs can be tested using the null
stream approach. This method is based on a mathematical fact that if GW has
pure tensor or pure vector polarization then there exists a linear
combination of the signals detected by three interferometers which gives
null \cite{Gurs89,Wen05,Chat06,Hagi18}. The null combination depends on the
GW propagation direction and orientation of the interferometer arms which
are accurately known for the GW170817 event, but is independent of the
unknown orientation of the orbital plane.

We calculated the null streams for the GW170817 event assuming pure tensor
and pure vector polarizations. The results for the null stream combinations
are%
\begin{equation}
Null_{\text{tensor}}=0.205H+0.388L+0.407V,  \label{nt}
\end{equation}%
\begin{equation}
Null_{\text{vector}}=0.540H+0.310L+0.150V,  \label{nv}
\end{equation}%
where $H$, $L$ and $V$ are the signals in the LIGO-Hanford, LIGO-Livingston
and Virgo detectors adjusted for the arrival time delays.

It is remarkable that very noisy Virgo signal enters the null stream for
vector GW (\ref{nv}) with a much smaller weight than that for the tensor
null stream. As a consequence, the vector null stream is substantially less
noisy than the tensor null stream. Therefore, use of the vector null stream
combination is favorable for distinguishing between pure vector and pure
tensor polarizations for the GW170817 event.

\begin{figure}[h]
\centering
\hspace*{-0.5cm} \includegraphics[width=8cm,angle=270,origin=c]{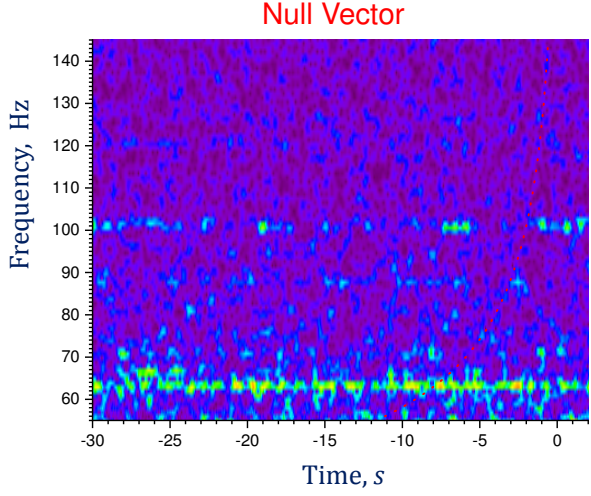}
\par
\vspace{-1.8cm}
\caption{Spectrogram of the null stream (\protect\ref{nv}) corresponding to
the vector polarization for the GW170817 event. The amplitude scale in the
spectrogram is the same as in Fig. \protect\ref{Fig2}. The red dotted line
indicates the expected position of the signal.}
\label{Fig20}
\end{figure}

In Fig. \ref{Fig20} we plot the spectrogram of the null stream corresponding
to the vector polarization for the GW170817 event. The spectrogram is not
very noisy, especially at high frequencies. No residual signal is visible in
the null-vector spectrogram which supports our conclusion about vector GW
polarization.

\begin{figure}[h]
\centering
\hspace*{-0.25cm} \includegraphics[width=7cm,angle=270,origin=c]{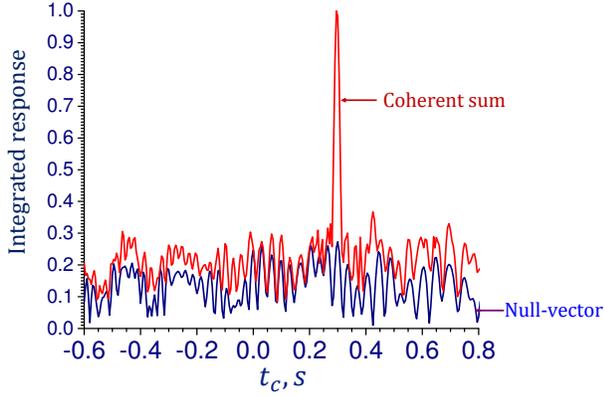}
\par
\vspace{-1.8cm}
\caption{Absolute value of the integrated response as a function of the
coalescence time $t_{c}$ for the coherent sum of the detector signals $I_{%
\text{max-v}}$ (red line) and the null stream for the vector GW polarization 
$I_{\text{null-v}}$ (blue line). }
\label{Fig21}
\end{figure}

By applying the signal accumulation approach to the null stream one can
obtain an upper-bound on the residual signal amplitude present in the vector
null stream. In Fig. \ref{Fig21} we plot the absolute value of the
integrated response as a function of the coalescence time $t_{c}$. The
signal is accumulated from \textquotedblleft good\textquotedblright\
frequency intervals $71-80$ Hz and $95-115$ Hz. The blue line shows the
integrated response for the vector null stream%
\begin{equation}
I_{\text{null-v}}=\left\vert I(0.540H+0.310L+0.150V)\right\vert .
\label{ivn}
\end{equation}%
It is at the level of noise for all $t_{c}$. The red line shows the sum of
the absolute values of the integrated responses for each detector taken with
the same weights as they enter the null stream%
\begin{equation}
I_{\text{max-v}}=0.540|I(H)|+0.310|I(L)|+0.150|I(V)|.
\end{equation}

The red line has a pronounced peak which yields the signal amplitude in the
strain data if the detector responses are added up constructively. One can
see from Fig. \ref{Fig21} that the amplitude of the residual signal in the
vector null stream (blue line) is compatible with zero.

Let us now assume that GW has tensor polarization. Then residual signal
should be present in the vector null stream and the ratio $I_{\text{null-v}%
}/I_{\text{max-v}}$ will be nonzero. The ratio depends on the inclination
and polarization angles $\theta $ and $\psi $. This dependence can be
calculated using Eqs. (\ref{wa5}), (\ref{s6}) and (\ref{s7}). In Fig. \ref%
{Fig22} we plot $I_{\text{null-v}}/I_{\text{max-v}}$ as a function of $%
\theta $ and $\psi $ for tensor GW. The red dashed area indicates values of $%
\theta $ compatible with the known distance to the source ($\theta <0.23\pi $
rad or $\theta >0.77\pi $ rad \cite{Abbo17c}). The plot shows that for the
range of the inclination angle compatible with the distance to the source
the ratio $I_{\text{null-v}}/I_{\text{max-v}}$ exceeds $0.38$. If so, for
tensor GW the residual signal should be visible in the blue curve $I_{\text{%
null-v}}$ of Fig. \ref{Fig21}, which is not the case. Hence, the vector null
stream combination, together with the additional constraint on $\theta $
based on the known distance to the source, rules out tensor GW polarization.

\begin{figure}[h]
\centering
\hspace*{-0.9cm} \includegraphics[width=7.7cm,angle=270,origin=c]{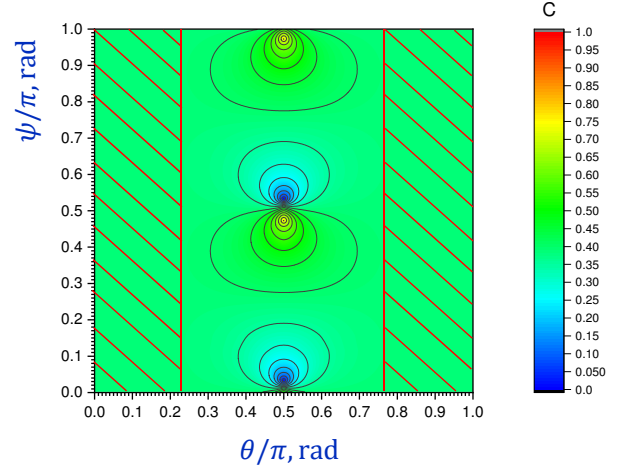}
\par
\vspace{-1.8cm}
\caption{Ratio of the vector null stream $I_{\text{null-v}}$ to $I_{\text{%
max-v}}$ for tensor GW for the GW170817 event as a function of the
inclination and polarization angles $\protect\theta $ and $\protect\psi $.
Red dashed area indicates values of $\protect\theta $ compatible with the
known distance to the source.}
\label{Fig22}
\end{figure}

\begin{figure}[h]
\centering
\hspace*{-0.3cm} \includegraphics[width=7cm,angle=270,origin=c]{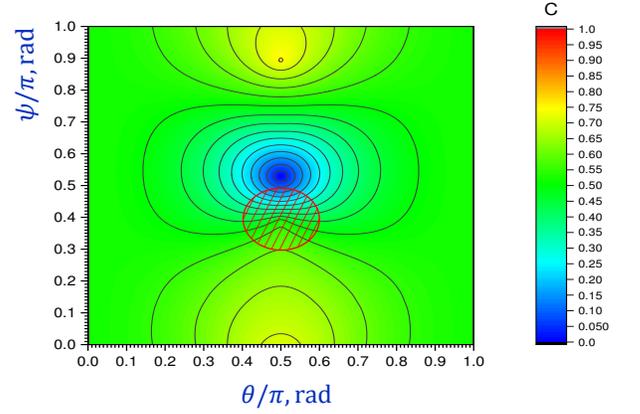}
\par
\vspace{-1.8cm}
\caption{Ratio of the tensor null stream $I_{\text{null-t}}$ to $I_{\text{%
max-t}}$ for vector GW for the GW170817 event as a function of the
inclination and polarization angles $\protect\theta $ and $\protect\psi $.
Red dashed area indicates values of $\protect\theta $ and $\protect\psi $
compatible with the known distance to the source.}
\label{Fig23}
\end{figure}

One should note that vector GW can produce a very small residual signal in
the tensor null stream (\ref{nt}) for the GW170817 event. In Fig. \ref{Fig23}
we plot $I_{\text{null-t}}/I_{\text{max-t}}$ as a function of $\theta $ and $%
\psi $ for vector GW. The plot shows that for certain values of $\theta $
and $\psi $ compatible with the known distance to the source (red dashed
area in Fig. \ref{Fig23}) the value of $I_{\text{null-t}}$ could be very
small. Thus, vector GW can mimic tensor polarization in the null stream
analysis of GW170817.

\section{Summary and Outlook}

The simultaneous detection of GWs by the three interferometers of the
LIGO-Virgo network, together with the known sky location of the source, can
be used to distinguish between general relativity \cite{Eins15} and vector
gravity \cite{Svid17,Svid18}, and rule out one of the two theories. However,
the possibility of coming to a decisive conclusion depends on how accurately
we can find the ratios of the signals ($H/L$ and $V/L$) measured by
different interferometers.

Here we have analyzed the data from GW170817 event produced by a pair of
inspiraling neutron stars and observed by the LIGO-Hanford, LIGO-Livingston
and Virgo interferometers. The optical counterpart of GW170817 was found
which yielded accurate localization of the source in close proximity to the
galaxy NGC 4993 \cite{Abbo17b}.

To obtain the $H/L$ and $V/L$ ratios, we used the strain time series for the
three detectors released by the LIGO-Virgo collaboration \cite{LV17}. For
the GW170817 event the GW signal is clearly visible in the LIGO-Livingston
and LIGO-Hanford data. However, this is not the case for the Virgo data due
to larger detector noise.

We extracted the signals from the noisy data by applying the signal
accumulation procedure, and then calculated $H/L$ and $V/L$ ratios and their
uncertainties. We found that signal ratios are consistent with the vector
theory of gravity \cite{Svid17,Svid18}. Also we found that vector gravity
yields a distance to the source in agreement with the astronomical
observations.

In contrast, we discovered that the signal ratios are inconsistent with
general relativity at the $2.5\sigma $ level. Moreover, we found that
general relativity is at odds with observations even if we use a much less
restrictive constraint based only on the $|H/L|$ ratio obtained directly
from the Fourier transform of the gravitational-wave strain data. If our
analysis is correct and the detectors were properly calibrated, Einstein's
general theory of relativity is ruled out at the $99\%$ confidence level and
future GW detections with three or more GW observatories should confirm this
conclusion with a greater accuracy.

One should mention that BAYESTAR \cite{Sing16} and LALInference \cite{Veit15}
codes commonly used for sky localization and estimation of the binary system
parameters tacitly assume that signal is not corrupted by noise filtering
and analyze the data from the entire detector bandwidth. In Section \ref%
{Comment} we showed that the measured LIGO-Livingston signal for GW170817 is
substantially reduced at certain frequency intervals which can be attributed
to noise filtering. We found that if these regions are excluded from the
analysis then data are consistent with vector GW polarization and not with
tensor. However, if the signal accumulation method is applied over the
entire detector bandwidth, including the regions in which the signal is
depleted by noise subtraction, the result underestimates the LIGO-Livingston
signal amplitude. That smaller amplitude then leads to an erroneous
conclusion that favors tensor polarization over vector polarization for the
GW. This is what the LIGO-Virgo collaboration claims \cite{Max18}.

What are possible alternatives to general relativity? Historically in the
literature, there have been many attempts at constructing different theories
of gravity and most of them were ruled out \cite{Will93,Will14}. To the best
of our knowledge, the only viable alternative theory, which also passes the
present test, is the vector theory of gravity \cite{Svid17,Svid18}. Despite
fundamental differences in the nature of the two theories, vector gravity
and general relativity are equivalent in the post-Newtonian limit. The two
theories also give the same quadrupole formula for the rate of energy loss
by orbiting binary stars due to emission of GWs.

In strong fields, vector gravity deviates substantially from general
relativity and yields no black holes. In particular, since the theory
predicts no event horizons, the end point of a gravitational collapse is not
a point singularity but rather a stable star with a reduced mass. We note
that black holes have never been observed directly and the usually-cited
evidence\ of their existence is based on the assumption that general
relativity provides the correct description of strong field gravitation.

In vector gravity, neutron stars can have substantially larger masses than
in general relativity and previous GW detection events can be interpreted in
the framework of vector gravity as produced by the inspiral of two neutron
stars rather than black holes \cite{Svid17}. Vector gravity predicts that
the upper mass limit for a nonrotating neutron star with a realistic
equation of state is of the order of $35$ M$_{\odot }$ (see Sec. $13$ in 
\cite{Svid17}). Stellar rotation can increase this limit to values in the
range of $50$ M$_{\odot }$. The predicted limit is consistent with masses of
compact objects discovered in X-ray binaries \cite{Casa14} and those
obtained from gravitational wave detections \cite{LIGO18}.

Vector gravity also predicts the existence of gaps in the neutron star mass
distribution, although the position of the gaps depends on the uncertain
equation of state. A $3-5M_{\odot }$ gap has been found in the low-mass part
of the measured compact object mass distribution in the Galaxy \cite%
{Ozel10,Farr11}. Vector gravity predicts that neutron stars with mass above
the $3-5M_{\odot }$ gap are very different from the low-mass counterpart
because they belong to a different branch of the star stability region and
have several orders of magnitude higher baryonic number density in their
interior.

Because properties of matter at such high density are unknown, the
composition of massive neutron stars is uncertain and could be very
different from the low-mass counterpart. A different composition might
result in a weaker emission of electromagnetic waves upon merger of these
objects. This could be a reason, in addition to a large distance to the
source, why optical counterparts were not discovered for GW events involving
massive objects. Another possible reason is that sky localization of the GW
sources was predicted incorrectly assuming tensor GW rather than vector, see
e.g. Fig. 6.2 in \cite{MaxPhD} for GW170814 sky location reconstructed under
the assumption of different polarization hypotheses. There are however some
indications that these types of GW mergers were actually followed by
electromagnetic emission \cite{Conn16,Stal17}.

For cosmology, vector gravity gives the same evolution of the Universe as
general relativity with a cosmological constant and zero spatial curvature.
However, vector gravity, as mentioned in the Introduction, provides an
explanation of dark energy as the energy associated with a longitudinal
gravitational field induced by the expansion of the Universe and predicts,
with no free parameters, the value of the cosmological constant which agrees
with observations \cite{Svid17,Svid18}.

Vector gravity, if confirmed, can also lead to a breakthrough in the problem
of dark matter. Namely, the theory predicts that compact objects with masses
greater than $10^{5}M_{\odot }$ found in galactic centers have a
non-baryonic origin and, thus, an as-yet-undiscovered dark matter particle
is a likely ingredient of their composition. As a result, observations of
such objects can allow us to ascertain the nature of dark matter.

It is interesting to note that properties of supermassive compact objects at
galactic centers can be explained quantitatively in the framework of vector
gravity assuming they are made of dark matter axions and the axion mass is
about $0.6$ meV (see Sec. 15 in \cite{Svid17} and Ref. \cite{Svid07}).
Namely, those objects are axion bubbles. The bubble mass is concentrated in
a thin interface between two degenerate vacuum states of the axion field. If
the bubble radius is large, surface tension tends to contract the bubble.
When the radius is small, for a bubble-like object vector gravity
effectively produces a large repulsive potential which forces the bubble to
expand. As a result, the bubble radius oscillates between two turning
points. The radius of the $4\times 10^{6}M_{\odot }$ axion bubble at the
center of the Milky Way is predicted to oscillate with a period of $20$ mins
between $1$ $R_{\odot }$ and $1$ astronomical unit ($215$ $R_{\odot }$) \cite%
{Svid07}.

This prediction has important implications for capturing the first image of
the shadow of the supermassive object at the center of the Milky Way with
the Event Horizon Telescope (EHT) \cite{Godd17}. Namely, because of size
oscillation the relatively low-mass bubble at the center of the Milky Way
produces a shadow by bending light rays from the background sources only
during short time intervals when the bubble size is smaller or of the order
of the gravitational radius $r_{g}=2GM/c^{2}=17$ $R_{\odot }$. Since typical
EHT image collection time is several hours, the time averaging yields a much
weaker shadow than that expected from a static black hole in general
relativity. In the time-averaged image, the shadow will probably be
invisible.

One should mention that first image of the Milky Way center with Atacama
Large Millimeter Array at $3.5$ mm wavelength has been reported recently
(see Fig. 5 in \cite{Issa19}). The resolution of the detection is only
slightly greater than the size of the black hole shadow. Still, a decrease
in the intensity of light toward the center should be visible. But the image
gets brighter (not dimmer) closer to the center. This agrees with the
predictions of vector gravity.

On the other hand, a much heavier axion bubble in the M87 galaxy ($M=4\times
10^{9}M_{\odot }$ \cite{Wals13}) does not expand substantially during
oscillations. The bubble in M87 has a mass close to the upper limit and,
according to vector gravity, its radius is close to $r_{g}/4$ \cite{Svid07}.
For such bubble the gravitational redshift of the bubble interior is $%
z=e^{2}-1\approx 6.4$. The redshift reduces radiation power coming from
matter trapped inside the bubble by a factor of $(1+z)^{2}=e^{4}\approx 55$,
that is the bubble interior mimics a black hole.

The large accretion disk surrounding the bubble in M87 produces radio
emission which is imaged by EHT. The size of the dark hole in the disc
(which has a radius of $\approx 2.6r_{g}$) is determined by the size of the
innermost stable circular orbit. For massive particles, the location of the
innermost stable circular orbit for the exponential metric of vector gravity
in the curvature coordinates is $3.17r_{g}$ \cite{Boon19} which is close to
that in Schwarzschild spacetime $3r_{g}$ \cite{Photo}. As a consequence, the
axion bubble in M87 produces an image similar to that of a black hole.
Recent imaging of the supermassive compact object at the center of M87 with
EHT at $1.3$ mm \cite{EHT19} is consistent with this expectation. Due to
bubble oscillations the shadow of the M87 central object might vary on a
timescale of a few days, which can be studied in the future EHT campaigns.

Presumably some dark matter bubbles harbor a fast spinning massive neutron
star or a magnetar which produces relativistic jet and occasional outbursts
of radiation when matter accretion accumulates critical density of hydrogen
on the stellar surface to ignite runaway hydrogen fusion reactions.

This work was supported by the Air Force Office of Scientific Research
(Award No. FA9550-18-1-0141), the Office of Naval Research (Award Nos.
N00014-16-1-3054 and N00014-16-1-2578) and the Robert A. Welch Foundation
(Award A-1261). This research has made use of data obtained from the LIGO
Open Science Center (https://losc.ligo.org), a service of LIGO Laboratory,
the LIGO Scientific Collaboration and the Virgo Collaboration.


\begin{thebibliography}{99}
\bibitem{Remark} To obtain this estimate we applied data whitening procedure
which assigns larger weight to frequency regions with better detector strain
sensitivity.
\bibitem{Photo} One should note that in the exponential metric of vector
gravity the radius of the photon sphere is $1.65r_{g}$ \cite{Boon19} which
is also not too far from the value in the Schwarzschild metric $1.5r_{g}$.
\end{thebibliography}
\end{document}